\documentclass[a4paper,fleqn]{cas-dc}

\usepackage[numbers]{natbib}
\usepackage{hyperref}
\usepackage{soul}
\def\tsc#1{\csdef{#1}{\textsc{\lowercase{#1}}\xspace}}
\tsc{WGM}
\tsc{QE}
\tsc{EP}
\tsc{PMS}
\tsc{BEC}
\tsc{DE}
\begin{document}
\def\floatpagepagefraction{1}
\def\textpagefraction{.001}
\shorttitle{Modeling Hadronic Interactions in Ultra-High-Energy Cosmic Rays within Astrophysical Environments: A Parametric Approach}
\shortauthors{A. Condorelli et~al.}

\title [mode = title]{Modeling Hadronic Interactions in Ultra-High-Energy Cosmic Rays within Astrophysical Environments: A Parametric Approach}                      
%\tnotemark[1,2]

%\tnotetext[1]{This document is the results of the research  project funded by the National Science Foundation.}

%\tnotetext[2]{The second title footnote which is a longer text matter to fill through the whole text width and overflow into another line in the footnotes area of the first page.}

\author[1,2,3]{Antonio Condorelli}[                        orcid=0000-0001-5681-0086]
\ead{antonio.condorelli@unina.it}

%\address[1]{, Street 129, 1043 NX Amsterdam, The Netherlands}
\affiliation[1]{organization={Dipartimento di Fisica “Ettore Pancini”, Università degli studi di Napoli “Federico II”},
                city={Napoli},
%               citysep={}, % Uncomment if no comma needed between city and postcode
                country={Italy}}
\affiliation[2]{organization={INFN - Sezione di Napoli, Complesso Univ. Monte S. Angelo},
                city={Napoli},
%               citysep={}, % Uncomment if no comma needed between city and postcode
                country={Italy}}
\affiliation[3]{organization={Université Paris-Saclay, CNRS/IN2P3, IJCLab, 91405},
                city={Orsay},
%               citysep={}, % Uncomment if no comma needed between city and postcode
                country={France}}
\author[4,5]{Sergio Petrera}[orcid= 0000-0002-6029-1255]

\affiliation[4]{organization={Gran Sasso Science Institute,Via Francesco Crispi 7},
                city={L'Aquila},
%               citysep={}, % Uncomment if no comma needed between city and postcode
                country={Italy}}
\affiliation[5]{organization={INFN Laboratori Nazionali del Gran Sasso, via G. Acitelli 22},
                city={Assergi (AQ)},
%               citysep={}, % Uncomment if no comma needed between city and postcode
                country={Italy}}
\begin{abstract}
Interactions of ultra-high energy cosmic-rays (UHECRs) accelerated in astrophysical environments have been shown to shape the energy production rate of nuclei escaping from the confinement zone.  
{To address the influence of hadronic interactions, Hadronic Interaction Models (HIMs) come into play. In this context, we present a parameterization capable of capturing the outcomes of two distinct HIMs, namely  EPOS-LHC and Sibyll2.3d, in terms of secondary fluxes, including escaping nuclei, nucleons, neutrinos, photons, and electrons.}
Our parametrization is systematically evaluated against the source codes, both at fixed energy and mass, as well as in a physical case scenario. The comparison demonstrates that our parameterization aligns well with the source codes, establishing its reliability as a viable alternative for analytical or fast Monte Carlo approaches dedicated to the study of UHECR propagation within source environments. This suggests the potential for utilizing our parameterization as a practical substitute in studies focused on the intricate dynamics of ultra-high energy cosmic rays.
\end{abstract}

%\begin{graphicalabstract}
%\includegraphics{figs/cas-grabs.pdf}
%\end{graphicalabstract}

%\begin{highlights}
%\item Research highlights item 1
%\item Research highlights item 2
%\item Research highlights item 3
%\end{highlights}

\begin{keywords}
Hadronic interactions \sep Source environment \sep Ultra-high-energy cosmic rays \sep Cosmic ray propagation \sep Multi-messenger astrophysics
\end{keywords}

\maketitle

\section{Introduction} \label{sec:intro}
%Although they have been known for more than a century, the nature of the ultra-high-energy cosmic-rays (UHECRs) sources still remains elusive.
The nature of the ultra-high-energy cosmic-rays (UHECRs) still remains elusive. Yet there are candidate astrophysical sources that appear to possess the necessary requirements to produce such energetic  particles \cite{Bustamante_2015, PhysRevD.97.063010, Fang_2018, Piran_2023, Peretti:2023uku}. In most cases they are extended sources where, apart the actual accelerators, the surrounding environment is filled with magnetic and radiation fields, as well as gaseous matter.

Cosmic-ray interactions in plausible source environments have proven to be a critical input for understanding the energy spectrum and mass-composition provided by Pierre Auger Observatory \cite{Auger_spectrum,PierreAuger:2023bfx} and Telescope Array data \cite{ABBASI2023102864,PhysRevD.99.022002}. To account for the progressive increase in energy of the mean logarithmic mass number $\langle \ln A\rangle$,  it is assumed that each individual nuclear component has the same magnetic rigidity at the sources. In addition, by assuming sources with an emission spectrum as $E^{-\gamma}$, the small mixture of elements detected in data can be replicated if the spectral index $\gamma \leq 1$ ~\cite{Aloisio:2013hya,Taylor:2015rla,PierreAuger:2016use}. Such hard spectral indices are surprising at first glance, as typical expectations from first-order Fermi shock acceleration predict $\gamma$ to be $\gtrsim 2$. It has been suggested that hard values for $\gamma$ may represent the effect of in-source interactions in changing the ejected spectra if the index of protons ejected from the sources is softer~\cite{Globus:2015xga,Unger:2015laa,Supanitsky:2018jje,Biehl:2017zlw,Condorelli:2022vfa, Muzio:2022bak}. It was also shown that such a requirement is consistent with the data in the energy range across the ankle feature~\cite{Luce:2022awd,PierreAuger:2022atd}.

Focusing on the interactions occurring in the environment surrounding the sources imposes studying the hadronic interactions in these environments. While negligible for extra-galactic propagation, hadronic interactions could play an important role in a source environment. In fact, in some extended sources (e.g. in the nucleus of Starbust Galaxies) where the high density of target ($\simeq 10^2 \  cm^{-3}$) and high value of magnetic field {($B \simeq 200 \ \mu \rm G$)} could significantly enhance their chance to happen. 

Hadronic interactions are a fundamental ingredient to study in-source propagation, because they significantly shape the shower, both in terms of disintegration of accelerated nuclei and in the secondary production of neutrinos, electrons and photons.
%and electromagnetic particles. 
As shown in~\cite{Condorelli:2022vfa}, they can 
%significantly 
contribute much more than photo-hadronic interactions in the expected neutrino flux from the source environment.\\

Cosmic ray propagation codes largely used  in UHECR community, such as SimProp \cite{Aloisio:2012wj} and CRPropa \cite{Armengaud_2007}, are planned to be updated with modules simulating hadronic interactions \cite{Condorelli:2023tat, Morejon:2023zbw}, to be suited to study in all aspects source environments.

Proton-proton and proton-nucleus interactions are normally taken into account by using {Monte Carlo (MC) codes based upon different Hadronic Interaction Models (HIMs). There are a few models available to simulate the hadronic interactions used in cosmic-ray physics: in particular, EPOS-LHC \cite{Pierog:2013ria}, QGSJetII-04 \cite{Ostapchenko:2010vb} and Sibyll 2.3d \cite{Riehn_2020} have been recently re-tuned after LHC data and are the most commonly used. 
%It has to be pointed out that their initial purpose was their use in the shower development induced by UHECRs in the atmosphere, to predict the depth of the shower maxima $X_\mathrm{max}$ and the muon flux at ground.

%If we neglect HIMs, estimating the secondary fluxes coming from hadronic interactions by using a simple astrophysical scenario is not an easy task: some theoretical works use the so-called "leading pion approximation" \cite{1990acr..book.....B}, where at each interaction only the most energetic pion (or three pions) are followed in the shower. \antonio{This assumption is widely used especially in the presence of very soft injection spectra, as it is assumed that higher energy interactions cannot contribute to the production of lower energy photons and neutrinos.} This is clearly a limitation if neutrino flux to be investigated is several order of magnitude less energetic that the maximum energy reached in the source environment.
The diversity of particles generated by HIM codes strongly hinders simplification into conventional analytical or parametric models. Yet, for the study of the production from cosmic rays of stable secondaries such as neutrinos, photons, and electrons, some  simplistic frameworks have been employed to establish connections between the energies of these particles and their parent protons. One such approximation revolves around the relationship between the energy of a proton and that of its resultant pions, resulting in assuming a fixed value for it ($\delta$-approximation, see e.g. \cite{2014PhRvD..90l3014K}).
%, where it's commonly posited that the typical energy of a pion amounts to roughly one-fifth of the energy of its parent proton, expressed as $E_{\pi} \approx E_p/5$. 
Further simplification often assumes 
fixed energy fractions for the photons, electrons and neutrinos stemming from the pion and muon decays.
%an even distribution of energy among the four particles stemming from the decays $\pi^+ \rightarrow \mu^+ \nu_{\mu}$ and $\mu^+ \rightarrow e^+ \nu_e \bar{\nu}{\mu}$ (and equivalently for $\pi^-$). This approximation leads to an estimate of $E{\nu} \approx E_{\pi}/4 \approx E_p/20$ for the energy of neutrinos and $E_{\gamma} \approx E_p/10$ for photons resulting from neutral pion decays into two photons ($\pi^0 \rightarrow \gamma \gamma$).
However, it is crucial to recognize the inherent oversimplification in these estimates, which provide only
rough insights alongside the assumption of a primary spectrum.
%Pions exhibit a spectrum of energies, and similarly, secondary particles originating from muon decays manifest continuous energy distributions. 
%Therefore, the neutrinos produced by the interaction of a nucleus with a specific energy are anticipated to showcase a significantly broad range of energies, emphasizing the nuanced complexity underlying particle interactions within HIM codes.

%Together with HIMs, some theoretical works in literature use a simplistic approach to establish a relationship between the energy of neutrinos and that of their parent protons. This involves that the typical energy of a pion is approximately one-fifth of the energy of the parent proton, denoted as $E_{\pi} \approx E_p/5$. Furthermore, it's often assumed that the energy is evenly distributed among the four particles resulting from the decays $\pi^+ \rightarrow \mu^+ \nu_{\mu}$ and $\mu^+ \rightarrow e^+ \nu_e \bar{\nu}_{\mu}$ (and similarly for $\pi^-$), leading to an approximation of $E_{\nu} \approx E_{\pi}/4 \approx E_p/20$. For neutral pions decaying into two photons ($\pi^0 \rightarrow \gamma \gamma$), a similar estimation yields $E_{\gamma} \approx E_p/10$. However, these estimates are overly simplistic: this is because pions possess a spectrum of energies, and likewise, neutrinos originating from muon decays exhibit continuous energy distributions. Consequently, neutrinos produced by the interaction of a nucleus with a specific energy are expected to display a considerably broad range of energies.

Other works aim for a more realistic approach, taking into consideration the  inclusive distribution of produced pions derived from HIM simulation codes and consequently also of their secondary products, i.e., neutrinos and photons \cite{Kelner:2006tc, 2014PhRvD..90l3014K, Roulet_2021,Koldobskiy:2021nld}. Nevertheless, these parameterizations are limited to proton primaries and mostly do not explore the highest-energies regime.
%, focusing in the energy range $[10 \ \rm GeV, 10 \  \rm PeV]$. 

In our study, we demonstrate that in the context of the propagation of UHECRs within an astrophysical source environment, its prevailing conditions permit the parameterization not only of secondary stable leptons and photons but also of the  other interaction products responsible for the nuclear cascade therein.
%In this work we show that within the propagation of UHECRs in the environment of astrophysical sources the site conditions allow to parameterize not only the secondary stable leptons and photons but also the other interaction products which are responsible of the nuclear cascade which develops inside it.
%Providing a simple and efficient way to simulate hadronic interactions at the highest energies is thus a timely topic. For this purpose we have developed a Monte-Carlo simulation code based on parametric functions describing  the key features crucial for in-source propagation. In the following, 
Starting from HIM simulations, we parameterize the most important quantities, namely the cross section, the secondary production and the nuclear fragmentation. This allows us to take into account the effects of the hadronic interactions in a broad energy range by using only a few parametric functions. This procedure is currently implemented for EPOS-LHC and Sibyll2.3d. In the future, further models will be possibly considered.
%(such as QGSJETII-04, etc..).\\

The paper is organized as follows: we introduce the HIMs we aim to study and detail the most important quantities for our study in Section~\ref{Sec:HIMs}; the parameterization of the most important quantities in an astrophysical scenario is described in Section~\ref{Sec:Parameterizations}; we compare the two models to the corresponding source codes, in order to validate the parametric models and estimate their accuracies and performances, in Section~\ref{Sec:Comparison}; we discuss the outcomes of our simulation in a sample in-source case, showing the impact of our assumptions, in Section~\ref{Sec:Example}. We finally draw our conclusions in Section~\ref{Sec:Conclusion}.

\section{HIMs}
\label{Sec:HIMs}
The advent of the LHC has brought about remarkable strides in particle physics and our comprehension of high-energy interactions. One of the key areas of focus has been the enhancement of Monte-Carlo generators, which are crucial for simulating particle interactions at extreme energies. The primary emphasis in this development has been on hadron production. For HIMs used in cosmic-ray physics, this has led to substantial advancements in the modeling of 
%forward 
hadrons, their interactions and decays, with the ultimate goal of understanding the behavior of forward hadrons within the extensive cascades of secondary particles. Even minor adjustments in the {baryon and resonance production} or the introduction of collective hadronization can have far-reaching consequences on observables, such as the generation of muons \cite{Pierog:2017nes}.

%In the domain of cosmic-ray air showers, HIMs like QGSJet, Sibyll, and EPOS have been designed to play a crucial role. The significance lies in understanding the behavior of forward hadrons within the extensive cascades of secondary particles. 
%This is very important because these high-energy cosmic-ray showers are responsible for the generation of various particles through interactions and decays.
%Hence, HIMs have a critical role in unraveling the mysteries of cosmic-ray air showers. 
%It is essential to emphasize that 

The propagation of UHECRs in a medium is determined by the following parameters: the resident time $\tau_\mathrm{res}$, i.e. the duration UHECRs spend within the medium; the interaction time $\tau_\mathrm{int}$, i.e. the time interval before an interaction occurs;  the decay time $\tau_\mathrm{dec}$ which is relevant for unstable particles.
These time scales apply to both the primary cosmic rays and any potential secondary particles generated.

The resident time is influenced by factors such as the size of the medium and of the presence of magnetic fields. The interaction time depends upon the cross-sections of the involved processes and the matter density. Decay time relies on the particle's mean lifetime and Lorentz factor.

In the atmospheric propagation of UHECRs, the crucial parameters are the hadronic interaction times and decay times. The decay time, spanning a wide range of possible values, holds particular significance, although the interaction time also varies substantially among different particles. Conversely, the resident time is predominantly dictated by the air column density, typically measured in hundreds of g cm$^{-2}$.  As a result, in the evolution of atmospheric cascades, forward hadrons play the primary role, with their interaction and decay times constantly competing, while only very forward nuclear fragments and nucleons actively participate.

The propagation regime within an astrophysical source displays significant variability as far as hadronic interactions are concerned. Depending on the source characteristics, it can range from a mere transport (or transparent) regime, if the resident time is considerably shorter than the interaction time ($\tau_\mathrm{res} \ll \tau_\mathrm{int}$), to a full cascade regime, if the resident time greatly exceeds the interaction time ($\tau_\mathrm{res} \gg \tau_\mathrm{int}$).
Even in scenarios where the cascade regime dominates, the crucial interaction timescales primarily concern nuclei and nucleons. In fact unstable hadrons in all sources have decay times substantially shorter than their interaction times ($\tau_\mathrm{dec} \ll \tau_\mathrm{int}$), thereby impeding their interaction with matter.

From all that it becomes evident that the role of HIMs in the propagation of UHECRs within astrophysical sources differs significantly
%from its role in
 with respect to extensive air showers. In source environments, a %distinct division 
 separation emerges between the contributions of nuclei and nucleons, primarily driving the cascade process, and that of unstable hadrons, which instead generate fluxes of neutrinos, photons, and electrons.
}

In order to model the dynamics of hadronic interactions and their effects in astrophysical sources, three features have been chosen to characterize hadronic interactions: the inelastic cross sections, the rapidity distributions of secondary particles, and the fragmentation of nuclei.\\
\begin{itemize}
    \item[-] \textit{p-p and {nucleus-p} cross sections}: these cross sections are crucial ingredients to compute the interaction time. The timescale for the hadronic interaction reads:
    \begin{equation}
    \tau_\mathrm{int} = \large( {n_{\rm ISM} \,   \sigma_\mathrm{inel} \, c \large)}^{-1} ,
    \label{eq:time_spal}
    \end{equation}
    where $n_{\rm ISM}$ is the  Interstellar medium (ISM)  gas density and $\sigma_\mathrm{inel}$ is the inelastic cross section for proton-proton or nucleus-proton interactions. In this work we assume that matter is composed only of protons or hydrogen molecules. 
     \item[-] \textit{Secondary production}: The multiplicity of secondary particles and their rapidity spectra in both proton-proton and nucleus-proton interactions 
     %refer to the number and variety of particles produced as a result of these interactions. They 
     affect primarily the %observational signatures of astrophysical sources,
      production of photons and leptons, originating from their decays. Only nucleons, among all hadrons, affect
     %such as gamma-rays and neutrinos, but also 
     the cascade evolution in the source medium. A good description of secondary particles is crucial for interpreting these observations.
    \item[-] \textit{Fragmentation of nuclei}:  After a nucleus in the ISM undergoes inelastic process, it may fragment into various nuclei and %particles. 
     nucleons. The study of these fragmentation processes is crucial in astrophysics, because they  affect the development of the cascade and, in a second instance, the spectra of secondary particles. 
\end{itemize}

%For this reason, in 
In the following we will provide a description of the approach used for the parameterization of these three features for the two different HIMs.\\

\section{Parameterizations}
\label{Sec:Parameterizations}

The software that emulates hadronic interactions described in this paper is made publicly available, as PARISH (PARametric simulation of In-Source Hadronic interactions)~\cite{PARISH}.
This availability ensures transparent access to all the  details concerning the parametric assumptions incorporated within the code. As a consequence, rather than providing in this paper an exhaustive account of the parametric functions, we focus on demonstrating the validation of key distributions. These distributions are compared with those generated by the original HIM simulation codes to illustrate the level of accuracy achieved for event features crucial for understanding the propagation in UHECR sources. This validation is presented in the next two sections.

%We present this validation in two main sections: firstly, by examining fixed energy and mass of the projectile nucleus (Sec. \ref{Sec:Comparison}), and secondly, by considering  a specific in-source propagation case (Sec. \ref{Sec:Example}).
\begin{figure}

	\includegraphics[width=\columnwidth]{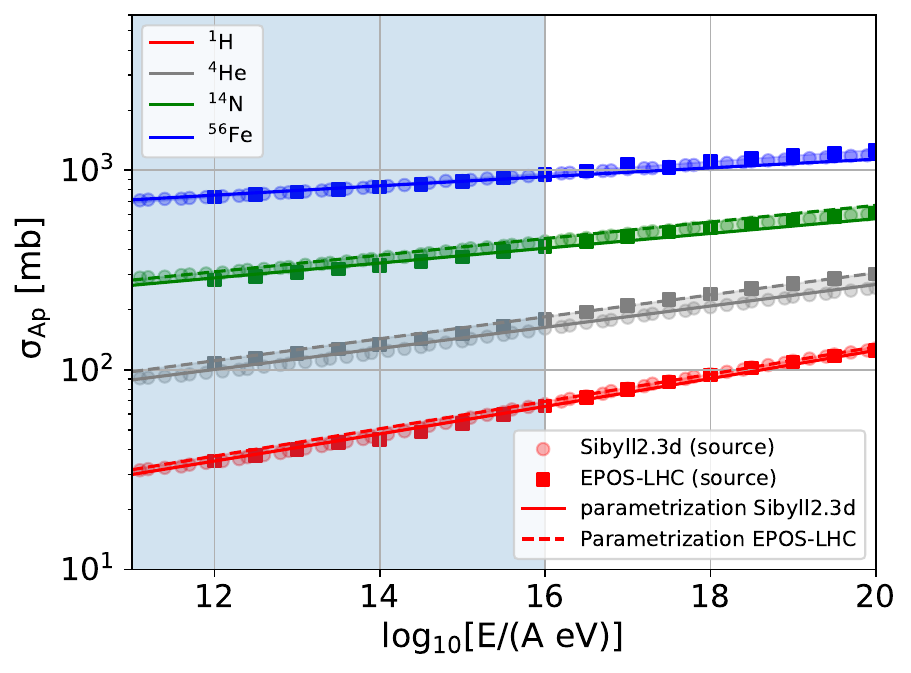}
    \caption{Inelastic cross-sections for nucleus-proton interactions for EPOS-LHC (solid) and Sibyll2.3d (dashed) lines for four nuclei ($^{1}$H, $^{4}$He, $^{14}$N and 
    $^{56}$Fe) in the energy-per-nucleon range $10^{11} \div 10^{20}$ eV/nucleon. {The data obtained by Sibyll2.3d and EPOS-LHC source code are also shown respectively with circles and squares. The shadowed blue area represents the energy region excluded from the fit procedure.}}

    \label{fig:CrossSection}
\end{figure}

\subsection{Inelastic cross sections}
\label{Ssec:ParXsect}

Nucleus-proton inelastic cross sections have been derived from EPOS-LHC  and Sibyll2.3d simulations.
They are easily fitted assuming logarithmic energy and mass dependence as shown in Fig. \ref{fig:CrossSection}. The fitted functions, though having slightly different parameters, {shows a good agreement with data. The fit was performed in the energy-per-nucleon range $10^{16} \div 10^{20}$ eV/nucleon, i.e. in the unshadowed area of Fig. \ref{fig:CrossSection}; in this region the residuals are of the order of  5\% for all the proposed nuclear masses. At lower energies the parameterization  describes the data to a satisfactory extent, with residual of 10\% at maximum. }
%Il fit e' fatto da 10^{16}, 
%nel range del fit i punti sono riprodotti con residui del 10%.
% Fuori dal range la param e' abbastanza fedele.
%Differenze fra epos & sibyll sono contenute entro il 10%
\subsection{Secondary particles}
\label{Ssec:ParHadrons}
HIM simulations produce a wide variety of hadrons with number multiplicities of the order of hundreds up to thousands, for energies in the UHE regime. As discussed in the previous section, the typical matter density in astrophysical sources is low enough to impede re-interactions of these particles, so all of them proceed through consecutive decays until only stable particles  survive (nucleons, photons, electrons and neutrinos). Nucleons\footnote{
Neutrons decay with a mean path of 9.2 $E$/EeV kpc. Whether they escape or decay depends on energy and source size. Generally, the chance of a neutron interacting as itself, rather than transforming into a proton post-decay, is very low.
} 
only can interact again in the source environment under certain conditions of matter density and diffusion, otherwise escape. Photons and electrons can induce further cascades in the radiation field depending on its strength, while neutrinos always escape. 

In these circumstances, pions and nucleons can stand as representative entities for all the hadrons generated in the interactions. This choice is supported by the following factors:
%In these conditions, pions and protons can serve as representatives for all the hadrons produced in the interactions. The reasons are the following:
\begin{itemize}
\item[a)] Pions account for more than 70\% of the total hadrons produced, while baryon-antibaryon pairs are produced at approximately a 10\% level.
\item[b)] Pions actively contribute to the production of electrons and neutrinos through charged pion-muon decays, as well as photons through neutral pion decays.
\item[c)] All baryons (antibaryons) heavier than nucleons (antinucleons) undergo successive decays until only nucleons (antinucleons) remain.
%\item[d)] nucleons and antinucleons exhibit the same behavior in successive interactions.
\end{itemize}

Therefore we  assume that all secondaries are either pions or nucleons and refer to it as ``lightest hadrons'' approximation. The two components serve distinct functions: the former contributes to the spectra of neutrinos, photons, and electrons, while the latter fuels the cascade process within the source medium.  

{The implementation of the `lightest hadrons' approximation is achieved by parameterizing the pion spectra to match those of the HIM source codes. For baryons, we assume they are instantly converted into nucleons upon generation, while preserving their original energy. Consequently, we fit the inclusive spectra of all baryons generated by HIM source code to represent those of the nucleons in our model. By parameterizing only pions and nucleons, we limit the decays we need to handle to those of pions and muons. This approach maintains computing times nearly constant, regardless of whether decays are enforced. }

{Although this method is simplistic and introduces some inaccuracies by not reproducing the decays of particles other than pions and muons, its impact is mitigated, as pions and nucleons constitute approximately 70\% and 10\% respectively of all produced particles. To estimate the effect of this simplification, we compare our results with the original HIMs, where all decays are enforced.
}

We derive the parametric distributions for pions and nucleons 
%(refer to the details below), 
utilizing simulated events from EPOS-LHC and Sibyll2.3d\footnote{
For Sibyll2.3d, the original source code was used to generate events. For EPOS-LHC, we used CRMC (Cosmic Ray Monte Carlo)~\cite{crmc}, an interface giving access to different generators. 
}. Our simulated data sample consists of nucleus-proton interactions in five logarithmic energy bins, log$_{10}$ ($E$/($A$ eV)) = \{16, 17, 18, 19, 20\} and four mass numbers $A$ = \{1, 4, 14, 56\}.
In each energy-mass bin, we have generated 5000 interactions for each HIM. These interactions include all hadrons that are produced instantly post-hadronization. Additionally, another sample of 5000 interactions per energy-mass bin has been generated, including only stable particles (post-decay), with the exception of neutrons, which are considered pre-decay. The former sample has been used to parameterize the inclusive distributions of pions and nucleons, the latter to validate the ones of neutrinos, electrons and photons. For the parameterization of nuclear fragmentation also samples of the same size have been produced for $^{28}$Si and $^{40}$Ca nuclei, to have a better description of the high-mass region.

In our parametric model, we focus on few key  observables, namely the multiplicity distribution of all secondaries and the rapidity distributions of pions and nucleons.  The distributions of neutrinos, electrons, and photons are not parameterized; rather, they are derived through MC simulation of the decay processes involving pions and muons. 

The approach to parameterization involves identifying a simple parametric form capable of reproducing the distribution from the source code across all energy-mass bins. Subsequently, we proceed determining the most straightforward evolution of the best-fit parameters. In the majority of cases, employing linear evolutions with respect to both the logarithms of energy-per-nucleon and mass number provides a satisfactory description across the entire energy-mass range.
When dealing with pion rapidity distributions in the context of EPOS-LHC, identifying a straightforward evolution of the parameters proved challenging. In this specific scenario, we resorted to a bilinear interpolation method to achieve an adequate representation.

Nucleon energy spectra are extracted from baryon and antibaryon rapidity distributions using HIM source codes.
%, operating under the assumption that these distributions are uniform across all baryonic particles.
  Unlike pions, their energy spectra do not demand an accurate modeling, as the cascade is solely influenced by the nucleon interaction length, which in turn exhibits a modest energy dependence. Consequently, we opt for a unified nucleon rapidity distribution with  identical parameters depending only on energy for both HIMs, overlooking the nuances between the two models.

\subsection{Nuclear fragmentation}
\label{Ssec:ParFrag}

Nuclear fragments, along with nucleons from secondaries (Sec. \ref{Ssec:ParHadrons}), contribute to the cascade within the source environment. Each nucleus interacts with ISM targets with comparable timescales (see eq.\eqref{eq:time_spal}), because inelastic cross sections  vary by at most a factor of 10  at a given energy-per-nucleon. 

Nuclear fragmentation is a slow process with respect to hadronic times, predominantly originated by the evaporation of a nucleus excited by the hadronic interaction (residual nucleus) \cite{PhysRevC.42.667,PhysRevC.28.950}. Upon examining the nuclei ejected in events from HIMs under consideration, we observed that  the inclusive mass distributions of produced fragments exhibit independence of the primary energy, aligning with expectations from the evaporation model. Consequently, we assumed the complete preservation of this energy-independence, and accordingly parameterized the fragmentation distributions solely as a function of the mass number, aggregating different primary energies present in the data sample.

Cosmic-ray collisions are predominantly peripheral, affecting only a small number of nucleons and imparting relatively modest energy to the residual nucleus (see e.g.~\cite{Engel:1992vf}). In addressing the subsequent de-excitation process, each HIM employs some statistical approach, involving the emission of nucleons or heavier fragments. In the context of UHECR interactions and in the source reference system, these fragments maintain the Lorentz factor of the initial projectile nucleus and follow its original direction.

The comparison between the outputs of the two HIMs reveals differences, hindering a uniform treatment of the parameterization process. In Sibyll2.3d, the nuclear fragments are distinctly categorized from the hadrons generated in the interaction. The fragments included in this class  encompasses nucleons and heavier nuclei. Consequently, the mass of the residual nucleus responsible for generating these fragments can be derived from the code output, along with the count of wounded nucleons\footnote{
The count of wounded nucleons is directly linked to the mass of the residual nucleus. The mass number of the residual nucleus may significantly differ from that of the projectile nucleus.
}. In contrast, the output of EPOS-LHC exclusively identifies nuclear fragments with $A>1$ via their $(A,Z)$ numbers, while nucleons originating from the evaporation remain untagged. 
Our parameterization assumes that, with $A$ representing the mass number of the projectile nucleus and $A_\mathrm{frag}$ the mass number of the nuclear fragment, the evaporation process involves also $A - A_\mathrm{frag} - 1$ nucleons.

Regarding the distribution of fragment masses, they have been parameterized for both HIMs. EPOS-LHC exhibits significant differences compared to Sibyll2.3d (see Sec. \ref{Sec:Comparison}). The former HIM generates a single heavy fragment ($A>1$), with rare exceptions when two $^{4}$He nuclei emerge from $^{8}$Be. All together, the produced fragments are too light with too many free nucleons, as pointed out in~\cite{Pierog:2023mee}. Sibyll2.3d displays a broad mass distribution of fragments, necessitating a more intricate parameterization,  based on iterative randomizations. This complexity arises because the determination of mass numbers for the fragments must consider the constraint imposed by the total mass number of the residual nucleus.

\begin{figure*}

	\includegraphics[width=\textwidth]	{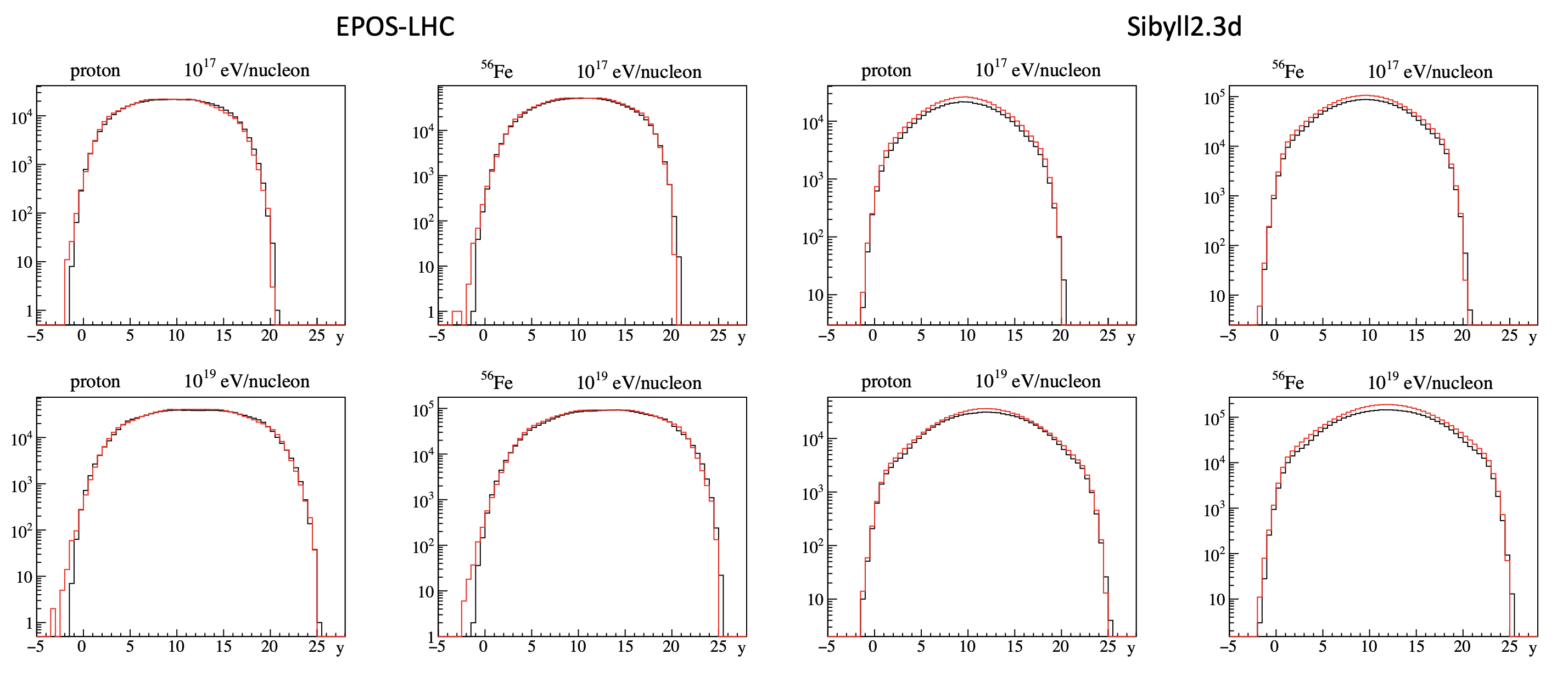}
    \caption{Pion rapidity distributions for protons and $^{56}$Fe nuclei of 10$^{17}$ and 10$^{19}$ eV/nucleon colliding with target protons. Left (right) panels correspond to EPOS-LHC (Sibyll2.3d). Black (red) line are generated with the corresponding HIM source code (parametric code) simulation.}

    \label{fig:yPion}
\end{figure*}
\begin{figure*}
	\includegraphics[width=\textwidth]	{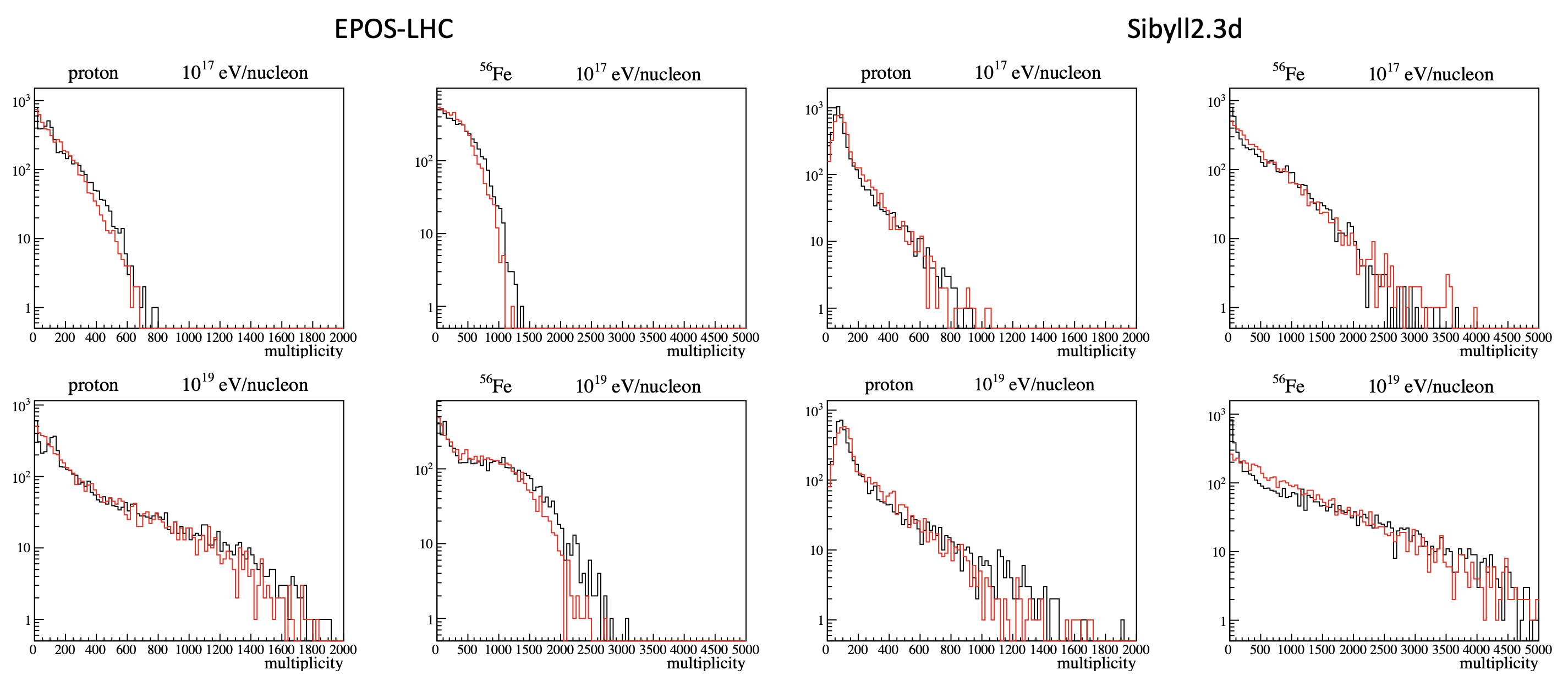}
    \caption{Distributions of total multiplicity for HIM source code (black) and {pion+nucleon} multiplicity for parametric model (red). Protons and $^{56}$Fe nuclei of 10$^{17}$ and 10$^{19}$ eV/nucleon colliding with target protons are shown. }

    \label{fig:multiplicity}
    
\end{figure*}

\section{Comparisons}
\label{Sec:Comparison}

In this section, we aim to demonstrate the validity of the parameterizations outlined earlier by conducting thorough comparisons between the distributions derived directly from the HIM source Monte Carlo codes and those generated by the corresponding parametric model. Additionally,  in the last part of the section, we provide comments about the usage of the parametric code and a comparative analysis with another established parameterization, focused on the production of secondary photons and neutrinos from protons.
%, which is further discussed towards the conclusion of this section.

\subsection{Secondary particles}
\label{Ssec:Secondaries}

Our parametric model is built upon the assumption of the `lightest hadrons' approximation. Concentrating solely on longitudinal kinematics, as extensively admitted in the context of UHECRs, the generation of secondary particles is derived exclusively from modeling the rapidity distribution of pions and nucleons. We have conducted a comparison between our distributions and those derived from HIM source simulations across each projectile energy-mass bin explored in this study. 
%For brevity, we present the comparison results at two energy-per-nucleon (10$^{17}$ and 10$^{19}$ eV/nucleon) for two projectiles: protons and $^{56}$Fe. 
%All the simulations (source and parametric codes) correspond to 5000 interactions in each energy-mass bin.

Figure \ref{fig:yPion}   shows the  inclusive pion rapidity distribution across four energy-mass bins. The black histograms represent data generated using the HIM source codes, while the red histograms depict results derived from the parametric model. The left set of panels corresponds to EPOS-LHC, whereas the right panels correspond to Sibyll2.3d. The shapes of pion rapidity distributions are reproduced to a satisfactory extent in the entire rapidity range and in all cases.  Parametric model tends to overestimate the HIM source rapidity distributions (by $10 \div 20\%$ for EPOS-LHC and $40 \div 50\%$ for Sibyll2.3d).
%, notably within the central region. 
This overestimation is necessitated by the partial inability of pion-muon decays to comprehensively represent the cascade decays of unaccounted particles.
 In fact, pions, being unobservable, are renormalised with
adjustment factors, to improve the agreement of secondary inclusive distributions.
%We notice that pions, being essentially unobservable in an astrophysical environment, do not require complete reproducibility. In contrast, neutrinos, electrons and photons are the only observable products originating from these pions.
%To address this, we have introduced adjustment factors in the pion-to-hadron and charged-to-neutral pion ratios, aiming to refine the agreement in the energy spectra of secondary stable particles.

%To elucidate this adjustment, 
Fig. \ref{fig:multiplicity} illustrates the distributions comparing the total multiplicity for HIM source codes (black) {against the multiplicity of pions+nucleons for the parametric model (red). Although these distributions are rather complex and significant differences emerge when comparing the two HIMs, the parameterizations effectively reproduce their main behavior. The mean multiplicities are within 10\% of the corresponding original HIM values, confirming the validity of the `lightest hadrons' approximation.}
%From these histograms we derive the ratio between the mean multiplicity of pions and all secondaries. Specifically, for EPOS-LHC, this ratio falls within the range of $0.74$ to $0.78$, while for Sibyll2.3d, it ranges between $0.90$ and 1. 
%These values denote an  increase compared to the ratios obtained directly from the HIM source codes, which approximate around $0.73$ (for all $A$) in the case of Sibyll2.3d, and gradually diminishing with $A$ from approximately $0.73$ to $0.69$ for EPOS-LHC.

%This adjustment toward higher values becomes necessary to increase the yield of neutrinos, electrons, and photons stemming from pion-muon decays, thereby partially compensating for the absence of certain hadrons not accounted for in the parametric model.

\begin{figure*}

	\includegraphics[width=\textwidth]	{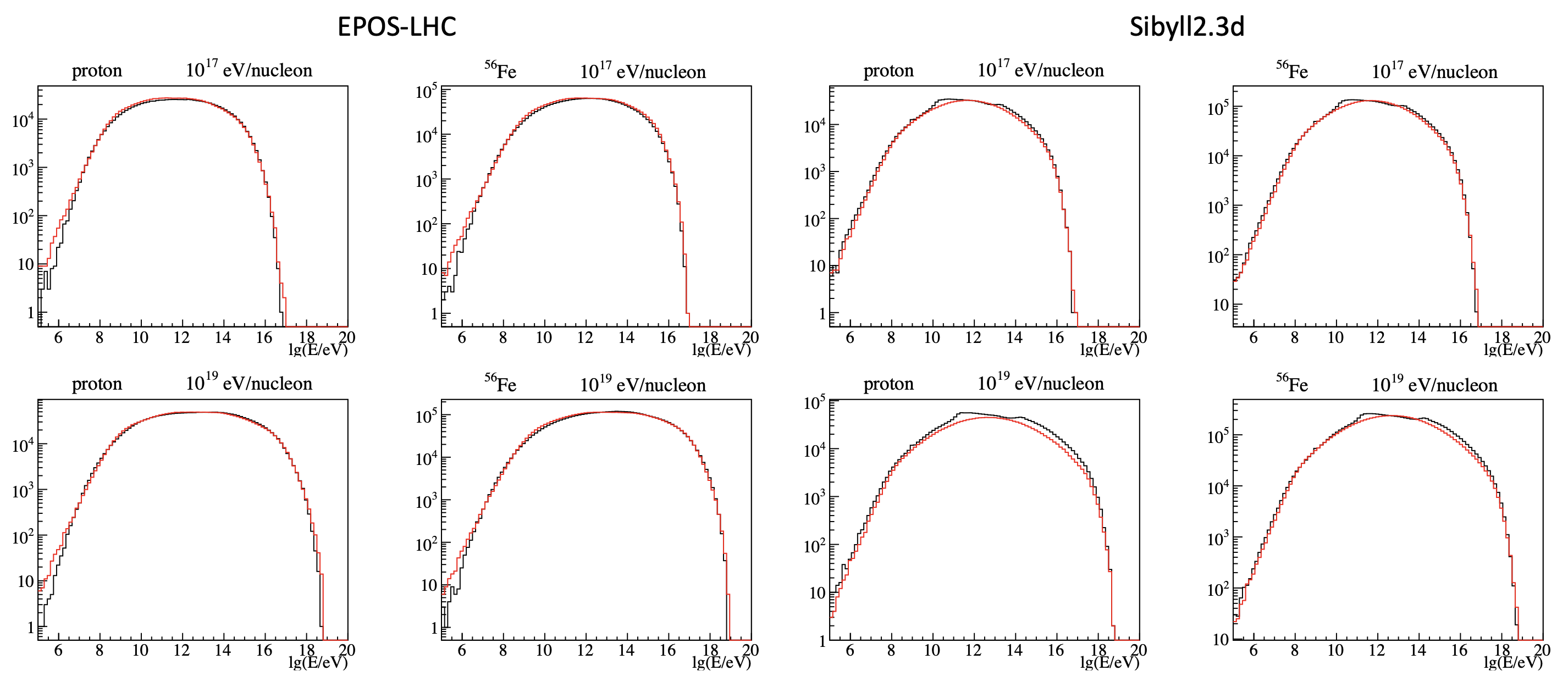}
    \caption{Neutrino log$_{10}(E$/eV) distributions for protons and $^{56}$Fe nuclei of 10$^{17}$ and 10$^{19}$ eV/nucleon colliding with target protons. Same color codes as in Fig. \ref{fig:yPion}.}

    \label{fig:EsecNu}
\end{figure*}

\begin{figure*}

	\includegraphics[width=\textwidth]	{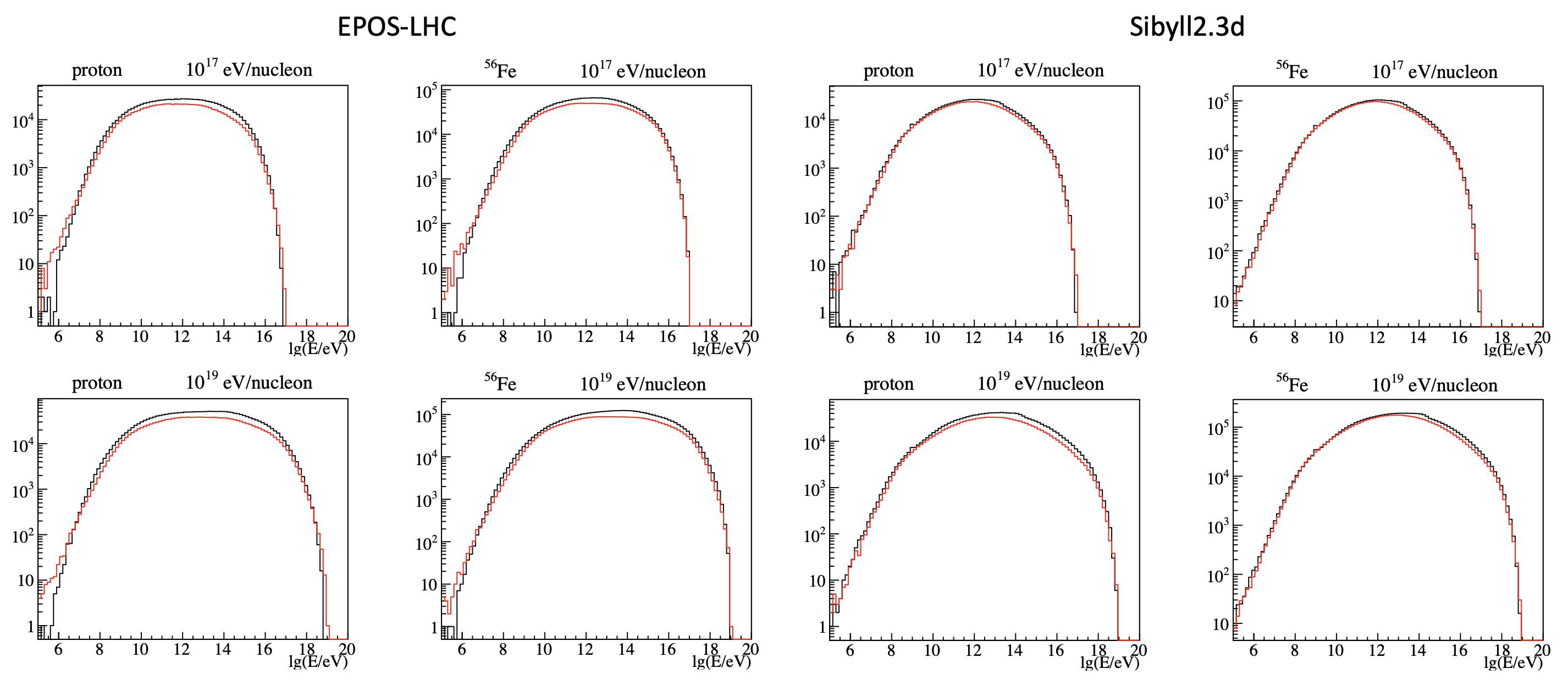}
    \caption{Log$_{10}(E$/eV) distributions of photons and electrons, for protons and $^{56}$Fe nuclei of 10$^{17}$ and 10$^{19}$ eV/nucleon colliding with target protons. Same color codes as in Fig. \ref{fig:yPion}.}

    \label{fig:EsecEM}
\end{figure*}

For each interacting nucleus of a given mass number and energy, the actual number $n_\pi$ of pions is randomly generated from the parametric multiplicity distribution. Subsequently, $n_\pi$ rapidity values are independently and randomly drawn from the corresponding rapidity distribution. Each rapidity value is converted into energy using the expression $E_\pi \simeq \mu_\pi/2 \times \exp{(y)}$, where $\mu_\pi \simeq$ 0.38 GeV represents the fixed assumed mean pion transverse  mass in the parametric model.

Having assumed that pions are independently drawn from inclusive rapidity distributions, energy-momentum conservation is not preserved on an event-by-event basis. Also correlations of pion energies with  multiplicity (as well as with the number of wounded nucleons, for nuclear primaries) are lost.
Despite this significant limitation, the observables of in-source propagation remain unaffected, provided the statistical sample is large enough (refer to Sec. \ref{Ssec:UsagePerf}).

Following the decay of pions, and the subsequent muon decays for charged pions, photons, neutrinos, and electrons are finally generated and the energy spectra of these secondary particles directly stem from the preceding distributions of pions. 
In the following, we will combine the energy spectra of electrons and photons, referring to them as `e.m. particles'. Despite originating from distinct interactions, they both progress together within the source medium through the electromagnetic cascade in the resident radiation field. The cascade process is not considered in this analysis, as it depends on source characteristics that fall outside the scope of this work.

Figures \ref{fig:EsecNu} and \ref{fig:EsecEM}
display the inclusive log-energy spectra of neutrinos and e.m. particles produced in interactions of 10$^{17}$ and 10$^{19}$ eV/nucleon protons and $^{56}$Fe nuclei with target protons (panels and colors. 
%Again the HIM source histograms are depicted in black and the parametric ones in red; the left panels refer to EPOS-LHC and the right ones to Sibyll2.3d. 
The reproduced neutrino spectra demonstrate a reasonable level of fidelity. In the case of EPOS-LHC, the parametric spectrum aligns within 20\% to the source spectrum, except for energies below about 1 GeV. However, for Sibyll2.3d, the parametric spectrum tends to underestimate the source spectrum by approximately 20\% to 40\%, particularly noticeable in certain energy intervals. 
Similarly, the energy spectra of e.m. particles are also reproduced to a reasonable extent, even if with a tendency for the parametric model to underestimate the HIM source spectrum. This underestimation amounts to about 40\% to 50\% for EPOS-LHC and 20\% to 40\% for Sibyll2.3d. One can argue that most of these disparities, for both neutrinos and e.m. particles, stem from hadrons whose decay modes and kinematics are poorly represented by pion-muon decays.

\begin{figure}\includegraphics[width=\columnwidth]	{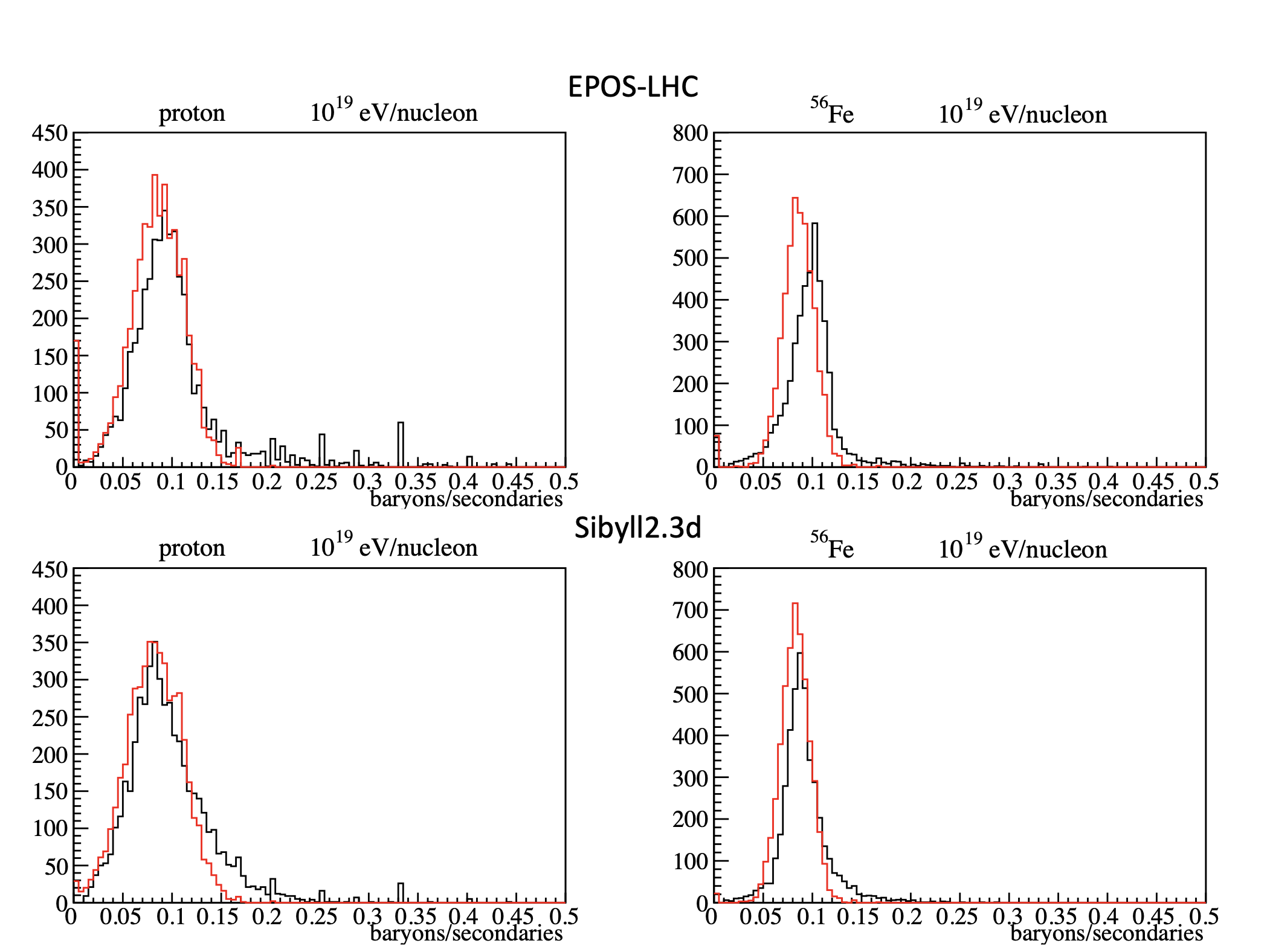}
    \caption{Distribution of the ratio of all baryons to the number of secondaries for protons and $^{56}$Fe at an energy of 10$^{19}$ eV/nucleon. Up (down) panel corresponds to EPOS-LHC (Sibyll2.3d). Black (red) line are generated with the corresponding HIM source code (parametric code) simulation.}

    \label{fig:fracB}
\end{figure}

\begin{figure}\includegraphics[width=0.9\columnwidth]	
{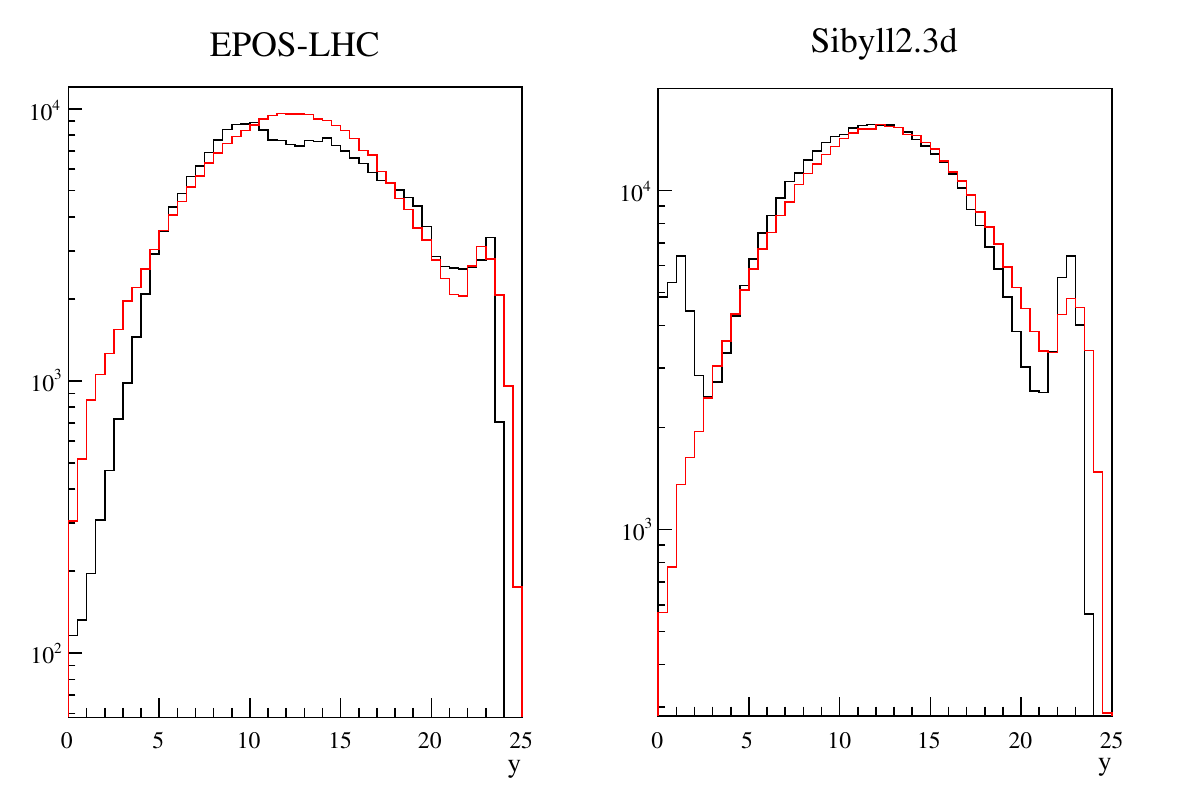}
\centering
    \caption{Rapidity distributions of baryonic particles, for $^{56}$Fe nuclei of 10$^{19}$ eV/nucleon colliding with target protons. Left (right) panel corresponds to EPOS-LHC (Sibyll2.3d). Same colors as for Fig. \ref{fig:fracB}.}

    \label{fig:yB}
\end{figure}

In relation to nucleons, their quantity is determined by modeling the ratio of all baryons to the number of secondaries, and studying its variation with the logarithmic energy-per-nucleon and mass number, as derived from the HIM source codes. This ratio is reasonably replicated across all energy-mass bins. In Figure \ref{fig:fracB}, the distributions of the baryonic fraction is presented for EPOS-LHC (above) and Sibyll2.3d (below), along with their respective parametric models, specifically for protons and $^{56}$Fe at an energy of 10$^{19}$ eV/nucleon.

Nucleon rapidity distributions have been studied for all the energy-mass bins.
As an example, Figure \ref{fig:yB} shows the rapidity distributions for $^{56}$Fe nuclei of 10$^{19}$ eV/nucleon colliding with target protons.  Modeling encompasses both the central region and the projectile leading peak. Instead, the target peak is excluded from consideration, given that the associated energy range falls significantly below the objectives outlined in this paper. It is important to note that the rapidity distributions include all baryonic particles in the HIM source code (depicted by the black histogram), while the parametric code represents the distribution for only protons or neutrons (in red), in accordance with the `lighest hadrons' approximation. Generally, the unified nucleon rapidity distribution assumed for both models better reproduces Sibyll2.3d than EPOS-LHC, which exhibits more complexity. However, the resulting differences in nucleon energy lead to minimal changes in the interaction length, as observed earlier.

\subsection{Nuclear fragmentation}
\label{Ssec:Fragmentation}

\begin{figure}

	\includegraphics[width=\columnwidth]	{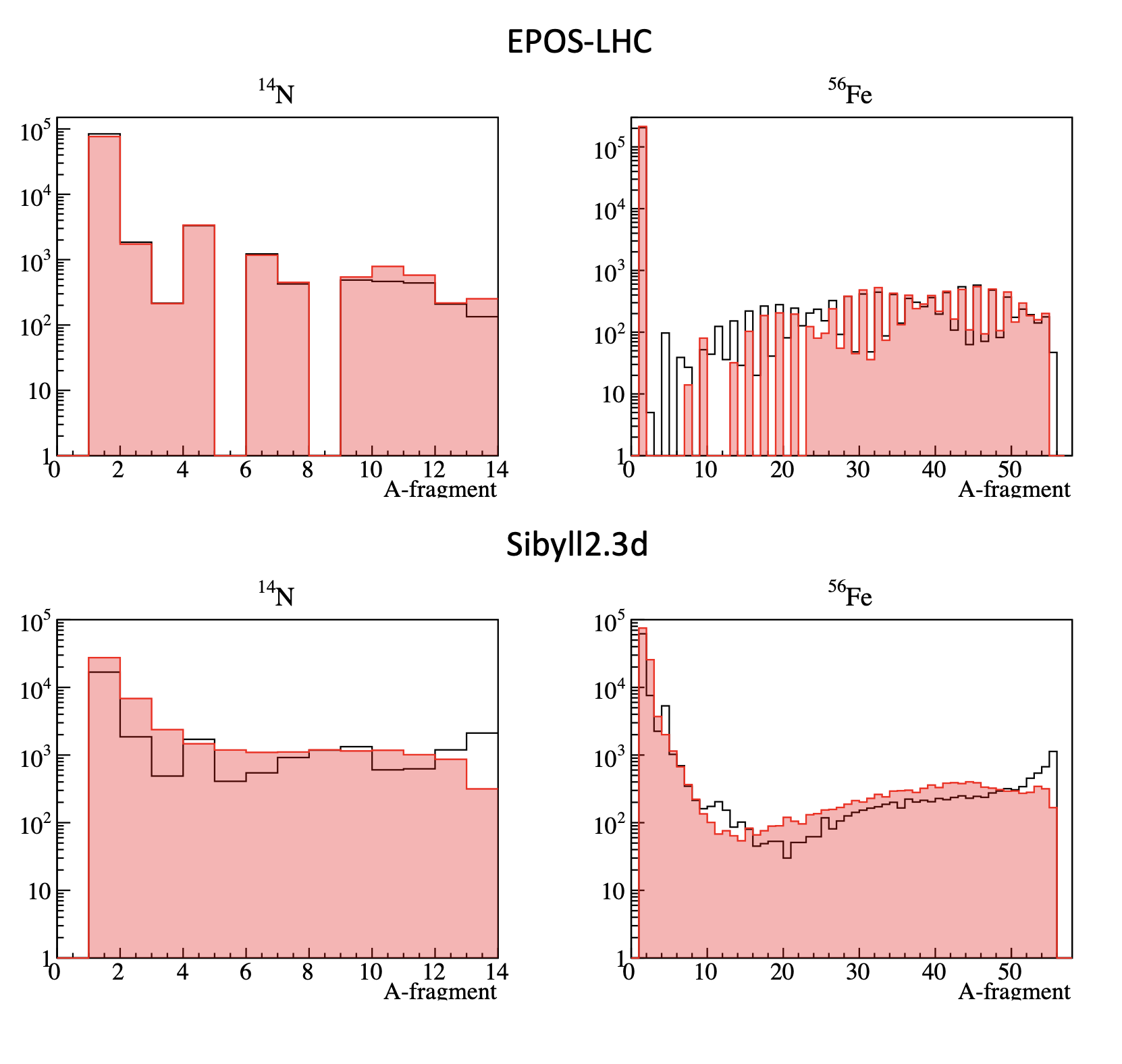}
    \caption{Distribution of the mass number of nuclei from the nuclear fragmentation of $^{14}$N and $^{56}$Fe. Up (down) panel corresponds to EPOS-LHC (Sibyll2.3d). Black (red) line are generated with the corresponding HIM source code (parametric code) simulation; parametric histograms are shaded for a better comparison with the source HIM ones.}

    \label{fig:Afrg}
\end{figure}

\begin{figure}

	\includegraphics[width=\columnwidth]	{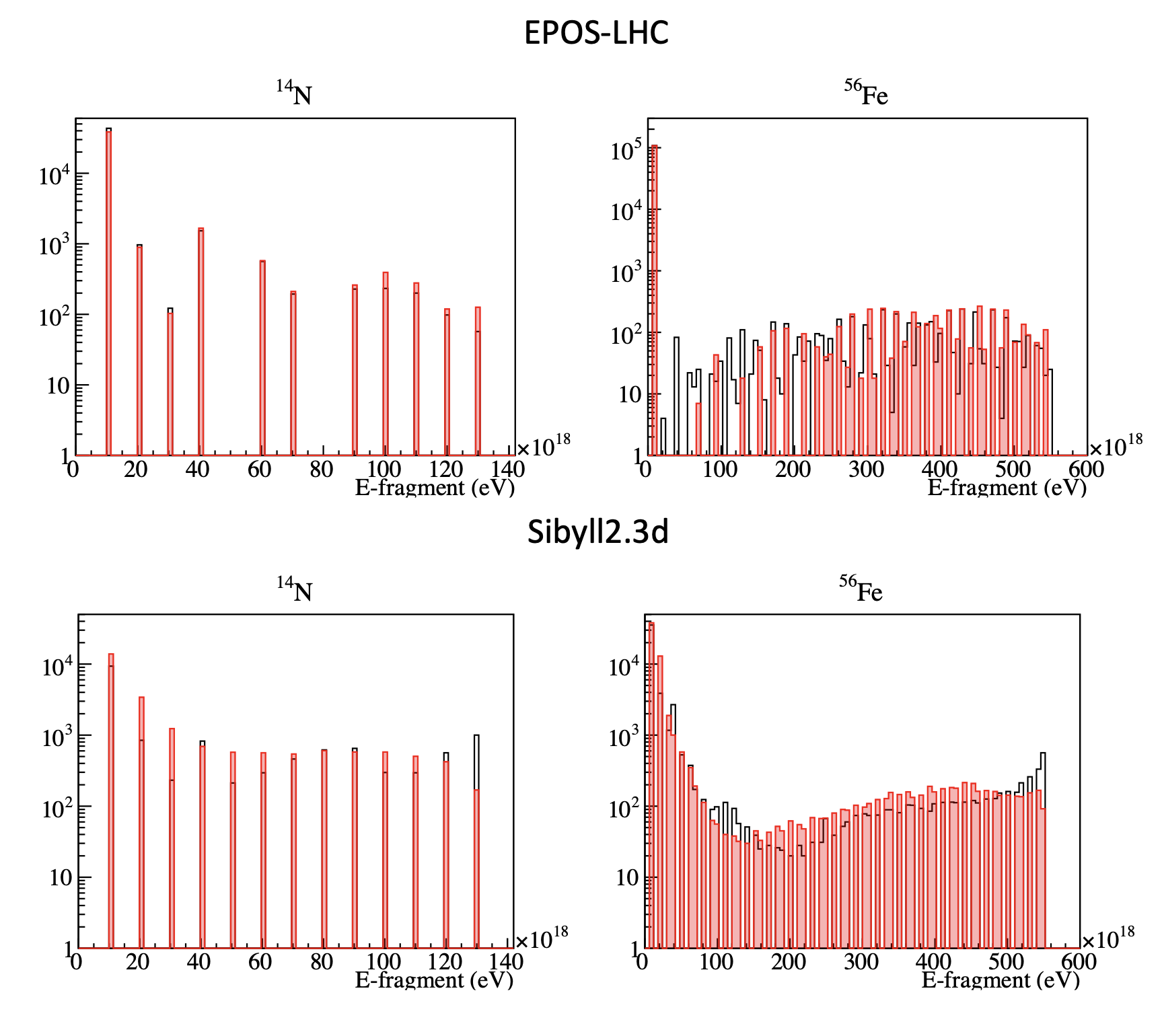}
    \caption{Energy spectra of the nuclear  fragments from $^{14}$N and $^{56}$Fe interactions at 10$^{19}$ eV/nucleon. Up (down) panels correspond to EPOS-LHC (Sibyll2.3d). Same colors as for Fig. \ref{fig:Afrg}.}

    \label{fig:EA}
\end{figure}

In Fig. \ref{fig:Afrg}, the distribution of nuclear mass numbers is shown for all fragments. The primary nuclei selected in this figure are $^{14}$N and $^{56}$Fe; the upper panels correspond to EPOS-LHC, while the lower panels represent Sibyll2.3d. The HIM source distributions are shown as black histograms, the parametric distributions are drawn as red shaded histograms.

%A notable observation is the 
We observe a significant disparity in the treatment of nuclear fragmentation between the two HIM source codes (black histograms). In EPOS-LHC, highly unstable nuclei within the lower mass range (e.g., 5 and 8) are absent from the list of fragments. Additionally, an evident odd-even mass number effect is observed. These distinct nuclear characteristics are presumably handled within the source code itself.
Conversely, Sibyll2.3d does not consider nuclear binding energy and stability, leading to the generation of all nuclei without discrimination, thereby leaving the treatment of unstable nuclei to the user. In our parameterization, we retained these individual features unchanged.

When comparing the outcomes of HIM simulations with those of the parametric model, we note the reproduction of the most significant features observed in the original HIMs. However, for Sibyll2.3d we notice some deficiency in mass numbers within the high mass range, particularly near the mass number of the projectile. This deficiency results in a tendency to favor lower mass numbers.
Regarding EPOS-LHC, while there is noticeable reproduction for $^{14}$N, difficulties manifest when attempting to produce lower mass numbers, in the case of $^{56}$Fe. All these disparities are expected to diminish their impact when considering propagation in astrophysical environments, as light nuclei are generated alongside heavier fragments to compose the final residual nucleus mass.
Given the comparable values of interaction mean paths among all nuclear fragments,  the main aspects of the cascading process are substantially preserved.

The energy spectra of the nuclear fragments exhibit a reasonable level of reproduction, as depicted in Fig. \ref{fig:EA}. This outcome is a direct consequence of preserving the Lorentz factor during the interaction. Then, the shape of the energy spectrum is determined by the distribution of fragment mass numbers.

\subsection{Usage and performances}
\label{Ssec:UsagePerf}

 The parametric simulation code (PARISH), described in this paper, is provided~\cite{PARISH} as a Monte Carlo generator for all particles and nuclei produced in single hadronic interactions. The collision products are listed as pairs  of particle IDs and energies with all particles treated longitudinally. In an astrophysical source environment, it is practical to enforce all hadronic decays so that only observable particles (nuclei, photons and leptons) are produced. 

As discussed in Sec. \ref{Ssec:ParHadrons}, the model assumes that hadronic secondaries are produced independently. Consequently, analyzing the products of a single interaction can easily lead to misinterpretations. Instead, the correct usage involves generating inclusive distributions of the interaction observables as elemental or secondary spectra. We have checked that these distributions remain unaffected by energy non-conservation, as the cumulative interactions number over few hundred.

We emphasize that the PARISH code is specifically tailored for integration into simulation frameworks within the context of extended astrophysical sources. 
%The primary function within these frameworks is to calculate the spectra of both nuclei and secondary particles that escape from the source. 
These simulation packages, such as CRPropa and SimProp, incorporate parametric models of photo-nuclear interactions. Consequently, our Monte Carlo simulation of hadronic interactions is ideally suited to complement and run concurrently with photo-nuclear interactions within these frameworks.

Efforts have already been made to integrate the original HIM source codes directly into propagation codes, either by interfacing with the HIM Fortran code itself  \cite{Condorelli:2022vfa}, by utilizing interaction tables generated through the CRMC interface \cite{Muzio:2021zud} or implementic a frontend interface to HIM's \cite{Morejon:2023zbw}. However, the adoption of a relatively simple C++ code, comprising around 1500 lines, as opposed to the original extensive source codes (approximately 25,000 and 82,000 lines for Sibyll and EPOS, respectively), facilitates straightforward insertion and management within the propagation codes. 

A relevant consideration is the CPU time required for generating hadronic interactions. For all the timing tests we utilized, as a reference, the same Linux machine with Intel(R) Core(TM) i7-7700 CPU @3.60GHz,  7.7GB of total memory. The system is running Ubuntu 18.04.6 LTS, { the installed C++ compiler is g++ (Ubuntu 7.5.0-3ubuntu1~18.04) 7.5.0 and the used Root version is 5.34/38.} The PARISH execution time per collision consistently remains below 6 ms (from 0.9 to 5 ms, for Sibyll2.3d, from protons to iron, $\approx$ 6 ms in all cases for EPOS-LHC). Remarkably, this value exhibits minimal variance across a broad energy range spanning from $10^{16}$ to $10^{20}$ eV/nucleon. Moreover, of significant importance is the observation that the execution time remains virtually identical for both the generation of secondary hadrons and secondary stable particles. 

The execution times for the same machine using the HIM source codes exhibit an increase with energy and projectile mass. Specifically, with Sibyll2.3d (EPOS-LHC), the duration per collision spans from 1 ms (18 ms) for protons at $10^{16}$ eV/nucleon to 9 ms (360 ms) for iron nuclei at $10^{20}$ eV/nucleon.

However, the most extensive utilization of HIM source codes involves their management of decays for all generated unstable hadrons. In this scenario, execution times per collision increase, ranging from 15 ms (22 ms) for protons at $10^{16}$ eV/nucleon to 41 ms (430 ms) for iron at $10^{20}$ eV/nucleon with Sibyll2.3d (EPOS-LHC).

The PARISH simulation, which includes particle decays, proves to be significantly more expedient than the source codes, with speeds ranging from approximately 10 to 70 times faster.

\begin{figure}

	\includegraphics[width=0.9\columnwidth]	{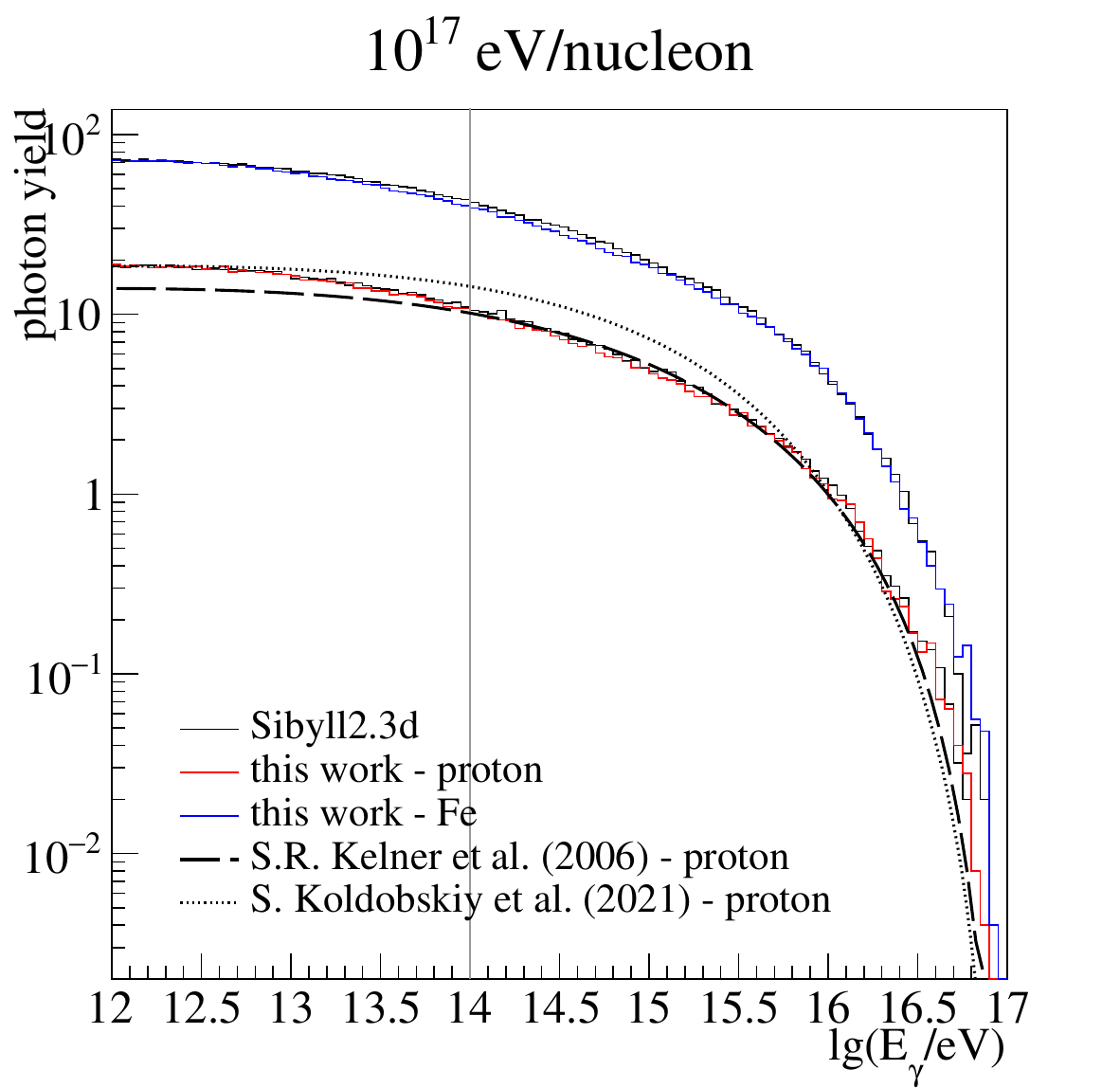}
    \caption{Differential photon yield per collision of $10^{17}$ eV/nucleon primaries. The black histograms represent photons generated using the Sibyll2.3d source code for proton (lower) and iron (upper) collisions. The red (blue) histogram corresponds to the Sibyll2.3d parameterization proposed in this study for protons (iron) collisions. The dashed line illustrates the analytical parametric function for photons {based on an older version of Sibyll,} as provided in \cite{Kelner:2006tc}: the vertical gray line marks the lower limit of applicability for this function. {The dotted line refers to the prediction in ref. \cite{Koldobskiy:2021nld}, based on QGSJetII-04m.}}

    \label{fig:photons}
\end{figure}

\subsection{Comparison with other models for secondary fluxes}
\label{Ssec:OtherModels}

 %In  extra-galactic cosmic ray propagation, hadronic interactions are commonly neglected due to the minimal matter density present in inter-galactic space. Conversely, in astrophysical sources, such interactions may occur, leading to the emergence of secondary fluxes, which can be of considerable interest. 
 In the field of High Energy cosmic ray physics, models concerning hadronic interactions primarily focus on the production of secondary neutrinos, electrons, and photons.

Simplistic approaches approximate the behavior of secondary leptons and photons as delta functions at a certain fraction of the primary energy. These offer only rudimentary insights alongside the assumption of a CR power-law differential spectrum, since pion multiplicity and the kinematics of their decays play an important role. 

The most widely accepted model is the one introduced by Kelner et al. \cite{Kelner:2006tc}.
The model provides simple analytical parametric functions for secondary photons and leptons, adaptable to any primary proton distribution. Only proton primaries are considered. 
Secondary particle spectra are characterized  as a function the ratio $x$ of energy transferred from the incident proton to the secondary particle, relative to the primary proton energy.  The model parameters are fitted using inclusive secondary meson spectra derived from Sibyll distributions, {(based on a pre-LHC version of this HIM)}, within the energy range of primary protons spanning from $10^{11}$ to $10^{17}$ eV, with the resulting secondary spectra being applicable for $x > 10^{-3}$. {More recent work has been reported by Koldobskiy et al. \cite{Koldobskiy:2021nld},  based on QGSJetII-04m, a post-LHC HIM not used in this work.} 

Figure \ref{fig:photons} illustrates a comparison of photon yields between the Sibyll parameterization proposed in this study and that formulated by Kelner et al. for primaries at $10^{17}$ eV/nucleon. {The parameterization by  Koldobskiy et al. is also shown in the same figure, though we do not use it for comparison, being based on a different HIM.} Additionally, the spectra from the Sibyll2.3d source code are depicted for both proton and iron collisions within the same figure. The photon distributions for protons from both Sibyll parameterizations align well with those generated by Sibyll2.3d and with each other, particularly for photon energies exceeding $10^{14}$ eV (i.e., $x > 10^{-3}$). Finally we highlight that our parameterization demonstrates remarkable fidelity in reproducing photon spectra also for iron nuclei.

\section{A physical case:\\ UHECR propagation in Ultra-luminous infrared galaxies}
\label{Sec:Example}

\begin{figure*}

	\includegraphics[width=\textwidth]	{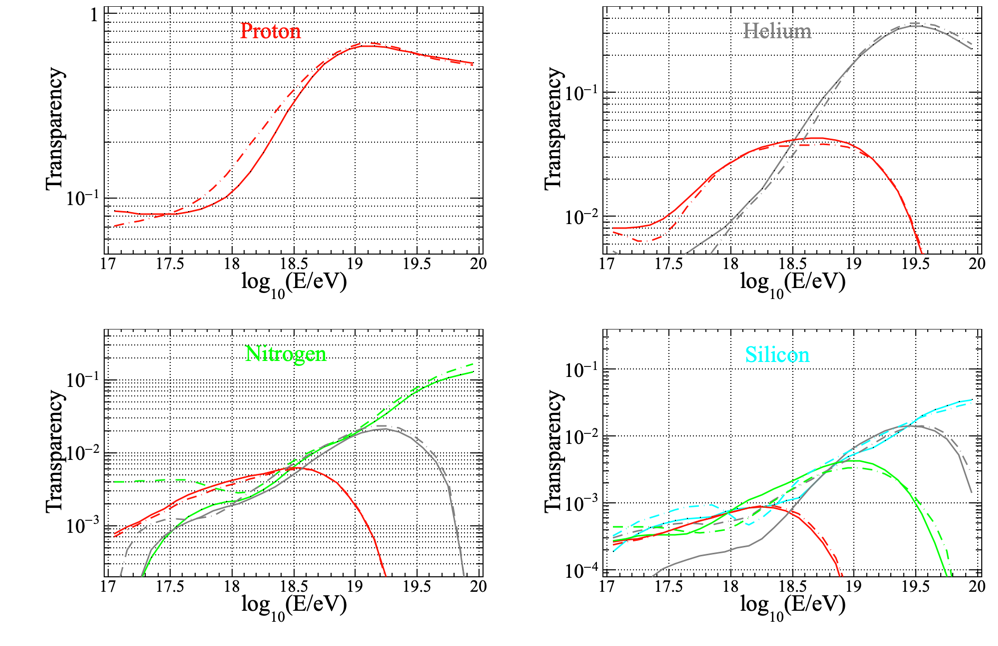}
    \caption{Escaping fluxes from the Starburst nucleus of Arp220 considering different injected nuclear species: protons (top-left panel), helium (top-right panel), nitrogen (bottom-left panel) and silicon (bottom-right panel). Dashed lines refer to Sibyll2.3d, solid lines to our parametric model. The escaping fluxes are normalized to the injection spectrum. The mass numbers of escaping fluxes are grouped as follows: $A = 1$ (red), $2 \leq A \leq 4$ (grey), $5 \leq A \leq 22$ (green), $23 \leq A \leq 28$ (cyan). The escaping fluxes are normalized to the injection spectrum.}

    \label{fig:Arp}
\end{figure*}

%The model parameters were derived through simulations that fixed the atomic number ($A$) of the nucleus and the energy-per-nucleon of the interaction, within the range of $10^{16}$ to $10^{20}$ eV/nucleon.
To assess the efficacy of the proposed parameterization, we give a concrete illustration where both the hadronic interaction model and the parameterization are employed in an astrophysical scenario. In this particular example, Sibyll2.3d   has been taken into consideration, utilizing SimProp as propagation code.

When UHECRs propagate within the source environment, a cascade involving nucleons and nuclear fragments is generated, whose evolution depends on various source features such as size, radiation, and magnetic fields.  Hadronic interactions occur at progressively lower energies, shaping the escaping nuclei, which form the observable energy spectrum and composition injected into the extra-galactic space.
Regarding the elemental energy spectra detected at UHECR observatories, the chosen energy range for parameterization is considered adequate for producing accurate model predictions. 

In contrast, secondary stable particles resulting from the same interactions possess energies at a per mil level compared to the colliding nucleus. 
Given the scientific importance of investigating the spectra of secondary neutrinos and electromagnetic particles, a key objective in multi-messenger astrophysics, we have carefully addressed the low-energy extrapolation of our model to maintain acceptable accuracy and robustness.
%The characteristics of these spectra are predominantly influenced by the multiplicity of secondary particles and the rapidity distribution of these particles. 
%The former is associated with the overall count of neutrinos and e.m. particles, whereas the latter determines the shape of the energy spectra. 
For this reason, the multiplicity of secondary particles and the rapidity distribution of these particles are tuned to best replicate the expected behavior according to the HIMs down to the 10$^{12}$ eV energy range.
 
A perfect environment where UHECR acceleration and high gas density could coexist is the core of ultra-luminous infrared galaxies ({ULIRGs}) \cite{ULIRG}. They are very dense environments, with a high rate of star formation and supernova explosions. They are thought to be sites of cosmic-ray acceleration, and are predicted to emit $\gamma$-rays in the GeV to TeV range. A high star formation rate means lots of young, massive, stars, which radiate mostly in the UV and have short lifespans on the order of tens of millions of years. The UV emission is absorbed by the interstellar dust and re-emitted in the far-infrared (FIR).

We focus in particular on  Arp220 \cite{Wiedner:2002fg} , the closest ULIRG to Earth, at a distance of about 75 kpc. It is the product of a merger of two galaxies and retains two dense nuclei, sites of very high star formation. Due to its high density and close proximity to Earth, it is considered a good candidate to search for high energy contribution from its star forming regions.
We choose to model the environment following \cite{Peretti:2018tmo}: a leaky box model is used, where interactions with photons and protons are taken into account, considering parameters at the source listed in Table \ref{tab:Reference_SBG}. The procedure adopted for this simulation is similar to \cite{Condorelli:2022vfa}.

\begin{table}
\begin{tabular}{ c|c } 
 \hline
{Parameter}  & {Value} \\ \hline
{$R \ \rm (pc)$}   & 250   \\
{$B \ \rm (\mu G)$}   & 500   \\
{$n_{\rm ISM} \ (\rm cm^{-3})$}   & 3500   \\
${U_{\rm eV \ cm^{-3}}^{\rm FIR} \ \bigg[\dfrac{kT}{\rm meV}\bigg] }$   &    31312 [3.5]  \\
${U^{\rm OPT}_{\rm eV \ cm^{-3}} \ \bigg[\dfrac{kT}{\rm meV}\bigg] }$   &    1566 [350]  \\
 \hline
\end{tabular}
 \caption{Parameters used to compute UHECRs propagation in the Arp220 environment (from \cite{Peretti:2018tmo}). The same notations of \cite{Condorelli:2022vfa} are used.}
    \label{tab:Reference_SBG}
\end{table}

In addition to considering interactions, it is essential to include diffusion in the environment; in fact, charged particles may remain confined for an extended period before eventually escaping. The diffusion timescale is given by $\tau_{\rm D}=R^2/D$, where $R \equiv pc/q \approx E/q$ represents the magnetic rigidity of a UHECR with energy $E$ and charge $q$. The diffusion coefficient, computed within the framework of quasi-linear theory \cite{Lee_2017}, is expressed as $D \simeq c r_{L}^{2-\delta} \, l_{\rm c}^{\delta -1} /3$. Here, $r_L = E/qB$ denotes the particle Larmor radius, $l_{\rm c}$ is the coherence length of the magnetic field, $\delta$ is the slope of the turbulence power spectrum, and $B$ represents the strength of the turbulent magnetic field. We adopt $\delta = 5/3$ as recommended for a Kolmogorov turbulence cascade.  

We assume  an energy spectrum $\Phi(E)$ following the diffusive shock acceleration with $\Phi(E) \propto E^{-2}$ and no cutoff.

We inject $10^4$ particles logarithmically distributed in the energy range $10^{17} - 10^{20} \ \rm eV$,  where our parameterizations are valid.
Note that the purpose of this exercise is not to test Arp220 as a possible source of UHECRs, but rather to estimate the effect of the surrounding environment on the propagation of UHECRs and thus to show that our parametric model describes escaping particles as well as HIM source code.

In Fig. \ref{fig:Arp} we show the transparency, i.e. the ratio between the escaping fluxes from the Arp220 environment and the injected ones, for different injected masses, from proton to silicon nuclei; dashed lines refer to Sibyll2.3d, solid lines to our parametric model. One can notice how the proposed parameterization can well describe the output of Sibyll2.3d in terms of secondary nuclei, 
with significant differences only at very low transparencies. These differences reflect in very small changes at the escape of the environment, meaning that the development of the cascade in the medium is accurately followed.
 One can also notice how the proton transparency present a maximum at energy $\simeq 10^{19} \  \rm eV$, which corresponds at the energy where the interplay between photo-hadronic and hadronic interaction times takes place.

\begin{figure}
	\includegraphics[width=\columnwidth]	{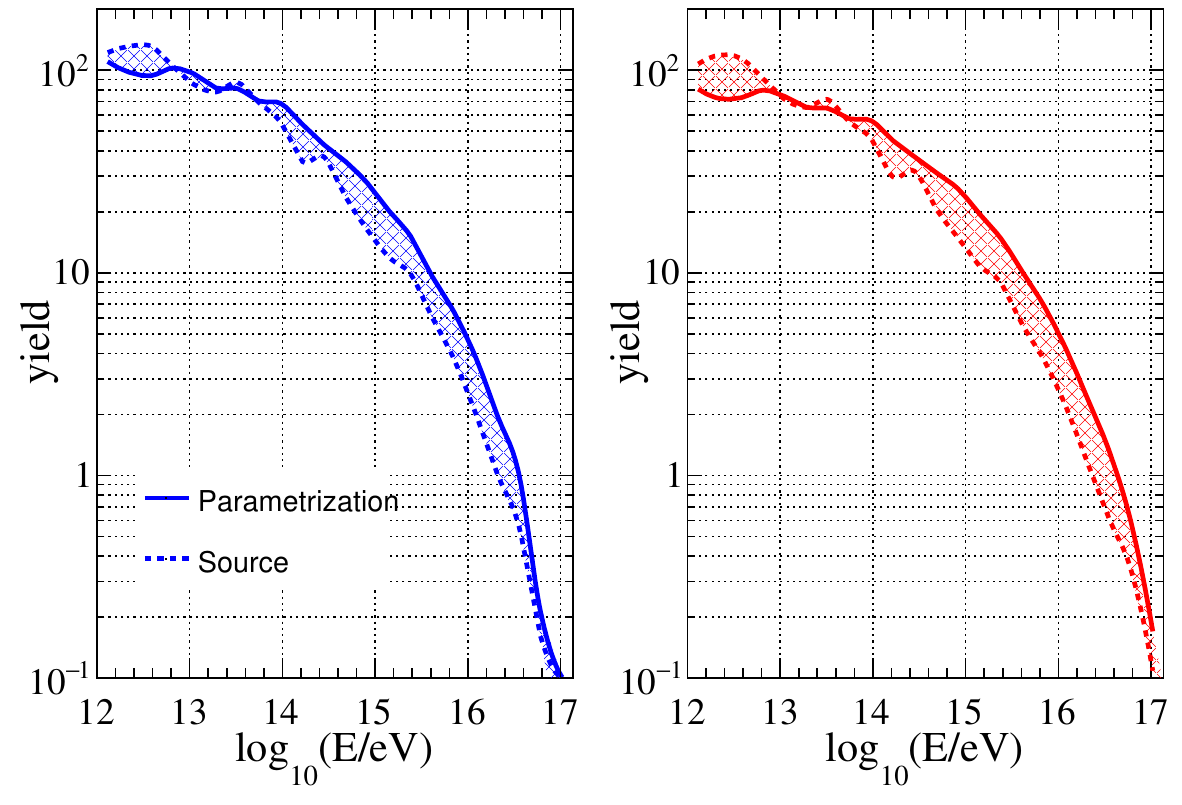}
    \caption{Neutrinos (left-hand side) and electromagnetic fluxes (right-hand side) associated to UHE proton propagation in Arp220. Dashed lines refer to Sibyll2.3d, solid lines to our parametric model. The escaping fluxes are normalized to the injection spectrum.}
    \label{fig:Neu}
\end{figure}
% per questa simulazione sono stati generati tot protoni da .. a ..
% cambia single

To test the secondary production, we extend our injection range down to $10^{12} \ \rm eV$, adopting the same number of primaries per energy bin as for nuclei. 
The production of neutrinos, photons and electrons in source environment are shown in Fig.\ref{fig:Neu}, where we compare the  yield, $ {\frac{dN_{i}}{dE}}/{\frac{dN_{\rm inj}}{dE}}$ i.e. the ratio of the number of secondary particles produced between $E$ and $E+dE$ to the number of protons injected in the same energy range.

The expected fluxes of neutrinos, photons and electrons produced in source environment are influenced by a dual interplay: firstly,  they are shaped by the development of the nuclear cascade, and secondly, by the production of secondary particles.
Also in this case a good agreement is reached, especially having in mind the uncertainties associated to the modeling of the source environment.
 It is crucial to highlight the distinct meanings of the two fluxes. In the case of neutrinos, the flux represents the actual neutrino spectrum emerging from the source environment. Conversely, for electromagnetic particles, determining the actual escaping flux necessitates incorporating the electromagnetic cascade in the source radiation field, a process not addressed in the current study.

\section{Conclusions} \label{Sec:Conclusion}
In this paper, we introduce a parameterization capable of describing the most important features of two different HIMs, Sibyll2.3d and EPOS-LHC, when hadronic interactions occur in an astrophysical source environment. This parameterization characterizes secondary fluxes, including disintegrated nuclei, neutrinos, photons, and electrons. Validation procedures demonstrate a good agreement between our parameterization and the underlying source code. Additionally, we illustrate this agreement within a physical scenario, showing how the combined effects of nucleus fragmentation and secondary production replicate cascades across a broad energy spectrum.

Our software, facilitating efficient emulation of the tested HIMs, is now publicly accessible, empowering the broader scientific community to enhance their predictions regarding hadronic interactions within source environments, without necessitating direct engagement with the HIMs themselves.

This study represents a significant advancement with respect to the previous literature; Introducing a dedicated parameterization for hadronic interactions is a novel approach for the UHECR community. We have shown that our proposed parameterizations offer competitiveness compared to direct HIM usage, with distinct advantages in terms of usability and computational efficiency.

Furthermore, there is potential for extending our parameterizations to lower energies, providing a versatile framework applicable to galactic cosmic rays as well.

The implications of this work extend to multi-messenger astrophysics, where these interactions play a pivotal role in shaping observed UHECR spectra. Accurately computing hadronic interactions within source environments is essential for precisely modeling UHECR energy spectra, compositions, and arrival directions. This, in turn, offers invaluable insights into the physical processes within the sources responsible for their acceleration.
\section*{Acknowledgments}
AC gratefully acknowledges funding from ANR via the grant MultI-messenger probe of Cosmic Ray Origins (MICRO), ANR-20-CE92-0052. The authors extend their appreciation to Francesco Salamida for valuable feedback that contributed to the development of the software. Special thanks to Denise Boncioli for her unwavering encouragement and support throughout the duration of this project.  Lastly, thanks go  to Felix Riehn  and Tanguy Pierog for their assistance in navigating the HIM codes.

\section*{Declaration of generative AI and AI-assisted technologies in the writing process.} 
Generative AI has been
used to improve text flow and grammar of existing text.

\bibliography{main}

\end{document}